\newcommand{\beq}{\begin{equation}}
\newcommand{\eeq}{\end{equation}}
\newcommand{\bea}{\begin{eqnarray}}
\newcommand{\eea}{\end{eqnarray}}

\newcommand{\gsim}{\lower.7ex\hbox{$\;\stackrel{\textstyle>}{\sim}\;$}}
\newcommand{\lsim}{\lower.7ex\hbox{$\;\stackrel{\textstyle<}{\sim}\;$}}

\def\simlt{\stackrel{<}{{}_\sim}}

\documentclass[aps,prl,twocolumn,preprintnumbers,%
               floatfix,nofootinbib]{revtex4}
\usepackage{graphicx}
\usepackage{mathrsfs}
\usepackage{amssymb}
\usepackage{verbatim}
\def\bea{\begin{eqnarray}}
\def\ea{\end{eqnarray}}

\def\stacksymbols #1#2#3#4{\def\theguybelow{#2}
    \def\vp{\lower#3pt}
    \def\sp{\baselineskip0pt\lineskip#4pt}
    \mathrel{\mathpalette\intermediary#1}}

\def\intermediary#1#2{\vp\vbox{\sp
     \everycr={}\tabskip0pt
     \halign{$\mathsurround0pt#1\hfil##\hfil$\crcr#2\crcr
              \theguybelow\crcr}}}

\def\comment#1{}

\def\u1x{U(1)_X}

\newcommand{\nc}{\newcommand}
\nc{\LL}{L} \nc{\vv}{\tilde{v}} \nc{\ccdot}{\!\cdot\!}
\nc{\gsm}{G_{SM}}
\nc{\vfive}{\mathbf{5}\oplus\mathbf{\overline{5}}}
\nc{\vten}{\mathbf{10}\oplus\mathbf{\overline{10}}}
\nc{\zhol}{Z^{\rm hol}}

\begin{document}

\preprint{~~~~IFT-10-04, DAMTP-2010-41, MIFPA-10-20}

\title{On Non-Canonical Kinetic Terms and the Tilt of the Power Spectrum}

\author{Sera Cremonini$^{a,b}$, Zygmunt Lalak$^c$ and Krzysztof Turzy\'nski$^c$}
\vspace{0.2cm}
\affiliation{
\vspace{0.1cm}
${}^{a}$Centre for Theoretical Cosmology, DAMTP, CMS,
University of Cambridge, Wilberforce Road, Cambridge, CB3 0WA, UK\\
${}^{b}$George and Cynthia Mitchell Institute for Fundamental Physics and Astronomy,
Texas A\&M University, College Station, TX 77843--4242, USA\\
${}^{c}$Institute of Theoretical Physics, Warsaw University, Ho\.za 69, 00-681, Warsaw, Poland
}





\begin{abstract}
We argue that in models of inflation with two scalar fields and non-canonical kinetic terms there is
a possibility of obtaining a red tilt of the power spectrum of curvature perturbations from
noncanonicality-induced interactions between the curvature and isocurvature perturbations.
We describe an extremely simple model realizing this idea, study numerically its predictions
for the perturbations and discuss applications in realistic scenarios of inflation.
We discuss to what extent in this model the scale of the inflationary potential can be decoupled from the
amplitude of the density fluctuations.
\end{abstract}

\maketitle

\setcounter{equation}{0}

By now inflation has become a paradigm of cosmological evolution, solving the horizon problem of the hot big bang scenario
and explaining the flatness of the universe. In addition, it provides a mechanism for
the origin of the $10^{-5}$ density contrast observed in the cosmic
microwave background and
the growth of large structure (see, e.g., \cite{mukhanov}).
However, it is still a subject of lively discussion how inflation fits into fundamental models of
particle physics, such as supergravity or, ultimately, string theory (see, e.g., \cite{Quevedo_review} for classical references
and \cite{mas_review} for a summary of recent developments).
In simple single-field models of slow-roll inflation, the measured normalization
of the power spectrum of the curvature perturbations $\mathcal{P}_\mathcal{R}(k)$
and its spectral index $n_s=1 +  \frac{\mathrm{d}\,\ln\mathcal{P}_\mathcal{R}}{\mathrm{d}\,\ln k}$
are given by:
\beq
\mathcal{P}_\mathcal{R} \simeq \frac{1}{24\pi^2M_P^4} \frac{V_\ast}{\epsilon_\ast}\, , \qquad n_s\simeq1-6\epsilon_\ast+2\eta_\ast\, .
\label{one}
\eeq
The slow-roll parameters are defined in terms of the Hubble expansion rate $H$
and the scalar field potential $V$ as
$\epsilon \equiv - \frac{\dot{H}}{H^2} \simeq \frac{M_P^2}{2} \bigl(\frac{V'}{V}\bigr)^2$,
$\eta\equiv \frac{V''}{3H^2}$.
The primes are derivatives with respect to the field $\chi$ that drives inflation,
and the subscript $\ast$ denotes the value at Hubble radius crossing.
With the observed deviation of $n_s$ from unity by few per cent \cite{Komatsu:2008hk}
and with a typical relation $\epsilon\simlt\eta$, one gets
$V^{1/4}\simlt 10^{16}\,\mathrm{GeV}$. In other words, observations require a hierarchy, $m\ll H\ll M_P$,
between the mass $m$ of the inflaton, the Hubble parameter and the Planck scale.
In the slow-roll approximation, the power spectrum can be rewritten as
$\mathcal{P}_\mathcal{R} \simeq \frac{1}{36\pi^2M_P^4}\frac{V^2_\ast}{\dot\chi^2_\ast}$.
Since it is natural to expect that with time the velocity of the field $\chi$ increases at the expense of the potential energy $V$,
the later a mode with a comoving wave number $k$ leaves the Hubble radius, the smaller power spectrum it corresponds to.
Hence, one expects a red-tilted power spectrum, $n_s<1$ \cite{mukhanov}. Similar conclusions hold for inflation accompanied by the evolution of another scalar field,
such as a modulus, driving the potential $V$ towards smaller values.

In this letter we propose a new mechanism for generating a red tilt of the curvature power spectrum
and we show that in specific realizations it can be the dominant one.
To this end, we go beyond minimal models of inflation and
consider scenarios with more than one scalar field active during inflation (see {\em e.g.} \cite{sy}),
focusing in particular on the role of non-canonical kinetic terms\footnote{For canonical kinetic terms, an enhanced red tilt can be arranged, {\em e.g.} as in \cite{Ashoorioon:2008qr}}.
This type of setting is well-motivated by known examples of string compactifications, which
typically contain a large number of light scalar fields $X^I$, or moduli, whose dynamics
is governed by a non-trivial moduli space metric $G_{IJ}$.
As long as the moduli-space metric is not flat,
one is generically led to non-canonical kinetic terms.
Although these effects may (but do not have to) be suppressed by the high scale of the corresponding
UV physics (e.g.\ moduli masses, string scale), they can still affect the inflationary dynamics
because of the above mentioned hierarchy of the mass parameters of inflationary models.

In this paper we will consider inflationary models described by an effective Lagrangian of the form
\cite{DiMarco:2002eb}:
\beq
\label{ourL}
\mathcal{L} = \frac{R}{16\pi G} +\frac{1}{2}\partial_\mu\phi \, \partial^\mu\phi
+\frac{e^{2\,b(\phi)}}{2}\, \partial_\mu\chi \, \partial^\mu\chi -V(\phi,\chi) \, .
\eeq
In many string motivated examples $M_P\,\frac{\mathrm{d}b}{\mathrm{d}\phi}$ can be rather large,
\emph{i.e.} of order 1.
The non-canonical kinetic terms can enhance the coupling between the curvature perturbations
(measured in the CMB) and the isocurvature perturbations, affecting the predictions of such models.
In this letter we show that such an \emph{enhanced coupling can result in a negative contribution to  the spectral index} of the curvature
perturbations. In a very simple model, we show that this can be the dominant effect responsible
for the deviation of the spectral index from unity.
 We defer a more complete and general treatment of the evolution of the curvature and isocurvature perturbations to \cite{Newpaper}, focusing here on a particularly simple case where the coupling is moderate.
Motivated by the plethora of scalar fields with non-canonical kinetic terms in effective
field theories originating from string theory,
we construct an analytically tractable toy
model of inflation, in which the noncanonicality
leads to interactions between curvature and isocurvature perturbations.

We begin with a brief review of the dynamics of cosmological perturbations during inflation.
The background equations of motion for (\ref{ourL}) are:
\bea
\label{bk1}
0 &=& \ddot{\phi}+3H\dot{\phi}+V_\phi-b_\phi e^{2b(\phi)}\dot\chi^2 \, ,\\
\label{bk3}
0 &=& \ddot{\chi}+3H\dot{\chi}+2b_\phi\dot{\phi}\,\dot{\chi}+e^{-2b(\phi)}V_\chi \, 
\eea
where $b_\phi\equiv \frac{\mathrm{d}b}{\mathrm{d}\phi}$. As usual, the Hubble parameter
is given by $H=(T+V)/(3M_P^2)$, where $T=(\dot\phi^2+e^{2b}\dot\chi^2)/2$.
Following \cite{Gordon:2000hv}, we utilize the so-called adiabatic-entropy decomposition, i.e.\
we introduce `polar coordinates' in the space of field velocities,
i.e.\ $\dot\sigma=(\dot\phi^2+e^{2b}\dot\chi^2)^{1/2}$, $\cos\theta=\dot\phi/\dot\sigma$ and
$\sin\theta=e^b\dot\chi/\dot\sigma$.
The slow-roll parameters
are then given by $\epsilon\equiv-\dot H/H^2$ and $\eta_{ij}\equiv \frac{V_{ij}}{3H^2}$,
where $V_{i_1,\ldots,i_n}$ denotes derivatives of the potential along directions
parallel ($\sigma$) or orthogonal ($s$) to the inflationary trajectory in the field space.
The instantaneous \emph{adiabatic} (curvature) and \emph{entropy}
(isocurvature) components read:
$Q_\sigma = \cos\theta Q_\phi +\sin\theta e^b Q_\chi$ and 
$\delta s = -\sin\theta Q_\phi+\cos\theta e^b Q_\chi$,
where $Q_\phi$ and $Q_\chi$ are the Mukhanov-Sasaki variables related to
the linear perturbations of the fields $\phi$ and $\chi$.
In the so-called comoving gauge, the perturbation $Q_\sigma$ is directly related to
the three-dimensional curvature $\mathcal{R}$ of the constant time slices by $\mathcal{R}=H(\dot\phi^2+e^{2b}\dot\chi^2)^{-1/2}Q_\sigma$.

The equations of motion for $Q_\sigma$ and $\delta s$ have been given in the closed form
in \cite{Lalak:2007vi}.  On scales larger than the Hubble radius and in the slow-roll case, $|\dot H|\ll H^2$, these
rather complicated formulae simplify to \cite{DiMarco:2005nq}:
\beq
\label{super2}
\frac{1}{H}\dot{Q}_\sigma = A\,{Q_\sigma}+B\,{\delta s} \, , \qquad \frac{1}{H}\dot{{\delta s}}=D\,{\delta s}\, ,
\eeq
where
\bea
\label{apar}
A&=&-\eta_{\sigma\sigma}+2\epsilon-\xi\cos\theta\sin2\theta\, ,\\
B&=& -2\eta_{\sigma s}+2\xi\sin3\theta \simeq \frac{2}{H}\dot{\theta}+2\xi\sin\theta\, , \\
D&=&-\eta_{ss}+\xi\cos\theta(1+\sin2\theta)
\label{dpar}
\eea
with $\xi = \sqrt{2\epsilon}b_\phi M_P$. From (\ref{super2}) it is apparent that the quantity $B$ parametrizes the \emph{coupling between the curvature and the isocurvature perturbations}.
Thanks to the presence of the noncanonicality, encoded by $\xi$,
this coupling does not vanish on super-Hubble scales even if $\dot{\theta}=0$.

Now we write down
a potential which allows for a large coupling between the
curvature and isocurvature perturbations. As follows from (\ref{super2})-(\ref{dpar}), the choice $\cos\theta=0$
makes the form of these equations identical to the case
of canonical kinetic terms -- with the exception of introducing a \emph{large
coupling} $\sim \mathcal{\sqrt{\epsilon}}$ between the curvature and the isocurvature modes.
Hence, such choice will make the impact of the noncanonicality on the inflationary spectra particularly
clear. Moreover, $\cos\theta=0$ corresponds to $\phi=\mathrm{const}$, which means that the normalization
of the inflaton field $\chi$, given by the prefactor $e^{2b(\phi)}$ in the kinetic term, remains unchanged
during inflation. Yet a nontrivial curvature of the field space metric enables the isocurvature perturbations
to affect the curvature perturbations even in the absence of a direct interaction term in the
potential, unlike in models in which the field space metric is flat.

From (\ref{bk1}) the requirement $\cos\theta=0$ corresponds to
$V_\phi=b_\phi e^{2b}\dot\chi^2$.
Although a trajectory $\cos\theta=0$ is a slow-roll ($\epsilon\ll 1$) solution to the full background
equations of motion (\ref{bk1})-(\ref{bk3}), it is {\em not} an approximate solution to the simplified
form of these equations with neglected double time derivatives.
What we have results from balancing the last two terms in (\ref{bk1}),
which are formally
of different order in the slow-roll parameters. Nevertheless, as we shall show in a numerical example, a moderate hierarchy between the parameters of the potential is sufficient for this purpose.
Since the usual slow roll solution $\dot\chi\approx M_P e^{-2b}V_\chi/(\sqrt{3V})$
is applicable here,
we find the following condition for a trajectory with $\cos\theta=0$:
\beq
\label{orel2}
3VV_\phi = M_P^2 b_\phi e^{-2b} V_\chi^2 \, .
\eeq
Note that the relation (\ref{orel2}) should be satisfied for a range of
values of $\chi$ relevant for inflation, but only for a single value of
$\phi$. Therefore, in order to look for appropriate evolution of the background fields we only need
to expand the potential $V$ to the linear order in $\phi$.
If the potential can be written as $V=U(\phi)\, \tilde U(\chi)$, we obtain from (\ref{orel2})
that $\tilde{U}'/\tilde{U}$ is a constant, which is solved by any function
$\tilde U$ proportional to $e^{\beta\, \chi/M_P}$ with an appropriate value of $\beta$.
In practice, it is enough to satisfy (\ref{orel2}) approximately,
e.g.\ by a function $\tilde U(\chi)$ admitting an expansion
$\tilde U_0(1+\beta(\chi-\chi_0)/M_P)$ over a sufficiently large range
of the arguments.

Following these observations, we shall from now on assume that the scalar potential
is very flat in {\em both} directions $\phi$ and $\chi$,
so that it suffices to expand $V$ as
\beq
V(\phi,\chi) = V_0\left(1+\alpha\cdot\frac{\phi-\phi_0}{M_P}+\beta \cdot\frac{\chi-\chi_0}{M_P}\right)\, ,
\label{exp1}
\eeq
and that the relations (\ref{orel2}) and $\dot\phi=0$ are satisfied as an initial condition.
As long as the potential can be reliably expanded as (\ref{exp1}), one
can estimate how much the field $\chi$ can shift during $N$ e-folds of inflation
without spoiling the condition $\dot\phi=0$. It follows from (\ref{bk1})
that this requires the field velocity $\dot\chi$ to be approximately constant,
which is guaranteed by the form of (\ref{exp1}).

Rewriting the equations (\ref{super2}) in terms of $\mathcal{R}=(H/{\dot\sigma})Q_\sigma$
and $\mathcal{S}=(H/{\dot\sigma})\delta s$
and noting that in our simple example the parameters $\tilde D=D-A$ and $B$ are constant,
one finds
\cite{DiMarco:2005nq} that
$
\mathcal{R}(N)=\mathcal{R}_\ast + (B/\tilde D)(e^{\tilde DN}-1)\mathcal{S}_\ast
$
and
$
\mathcal{S}(N)=e^{\tilde DN}\mathcal{S}_\ast
$,
where $N=\ln(a/a_\ast)$ is the number of efolds after the Hubble radius exit of the mode
with the smallest $k$.
Barring initial correlations and assuming that $|\tilde D| \, N\ll1$,
the power spectra at the end of inflation are given by:
\beq
\label{ana}
\mathcal{P}_\mathcal{R} = \mathcal{P}_\mathcal{R}^0\left(1+B_\ast^2N^2_\mathrm{e}\right) \, ,\,\,
\mathcal{P}_\mathcal{S} = \mathcal{P}_\mathcal{R}^0\left(1+2\tilde D_\ast
N_\mathrm{e}\right) \, ,
\eeq
where $N_\mathrm{e}$ denotes the number of e-folds between horizon crossing and the end of inflation.
The initial condition
$\mathcal{P}_\mathcal{R}^0=\frac{V_\ast}{24\pi^2\epsilon_\ast}$
for {\em both} the curvature and isocurvature perturbations
corresponds to the value that the power spectrum of each perturbation would have
in the absence of the coupling $B$.
Noting that $\mathrm{d}\ln k=\mathrm{d}\ln (a/a_\mathrm{e})=-\mathrm{d}N_\mathrm{e}$,
we arrive at the following expression for the running of the
spectral index of the curvature perturbations:
\beq
n_s-1 = -\frac{2B_\ast}{\frac{1}{B_\ast N_\mathrm{e}}+B_\ast N_\mathrm{e}}+\mathcal{O}(\epsilon_\ast)
\label{indan}
\eeq
(since the potential is linear, we have $\eta_{i,j}=0$ for $i,j=\sigma,s$ and
we consider a very flat potential).
The fraction in (\ref{indan}) is maximized for $BN_\mathrm{e}\sim 1$, in which case it
assumes a value $\sim 1/N_\mathrm{e}$.
Note, however, that during $N_\mathrm{e}$ efolds the field variation is
$\Delta\chi\approx \beta N_\mathrm{end}M_P$, while
$B\approx2\beta b_\phi M_P$, so the maximal effect is obtained for a field
variation close to the Planck scale.
The possibility of obtaining such a large field variation in string models
is a subject of intensive research (see {\em e.g.} \cite{msw}).

\begin{figure}[t]
\includegraphics*[height=5.5cm]{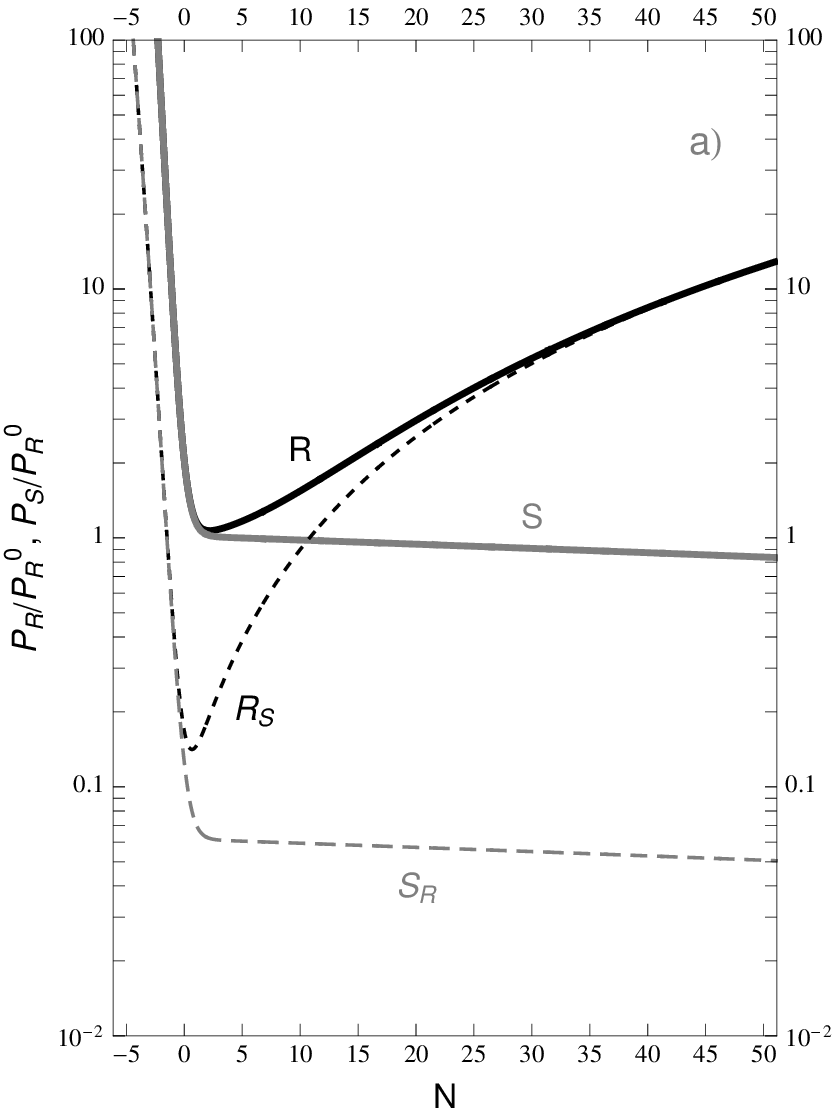}
\includegraphics*[height=5.5cm]{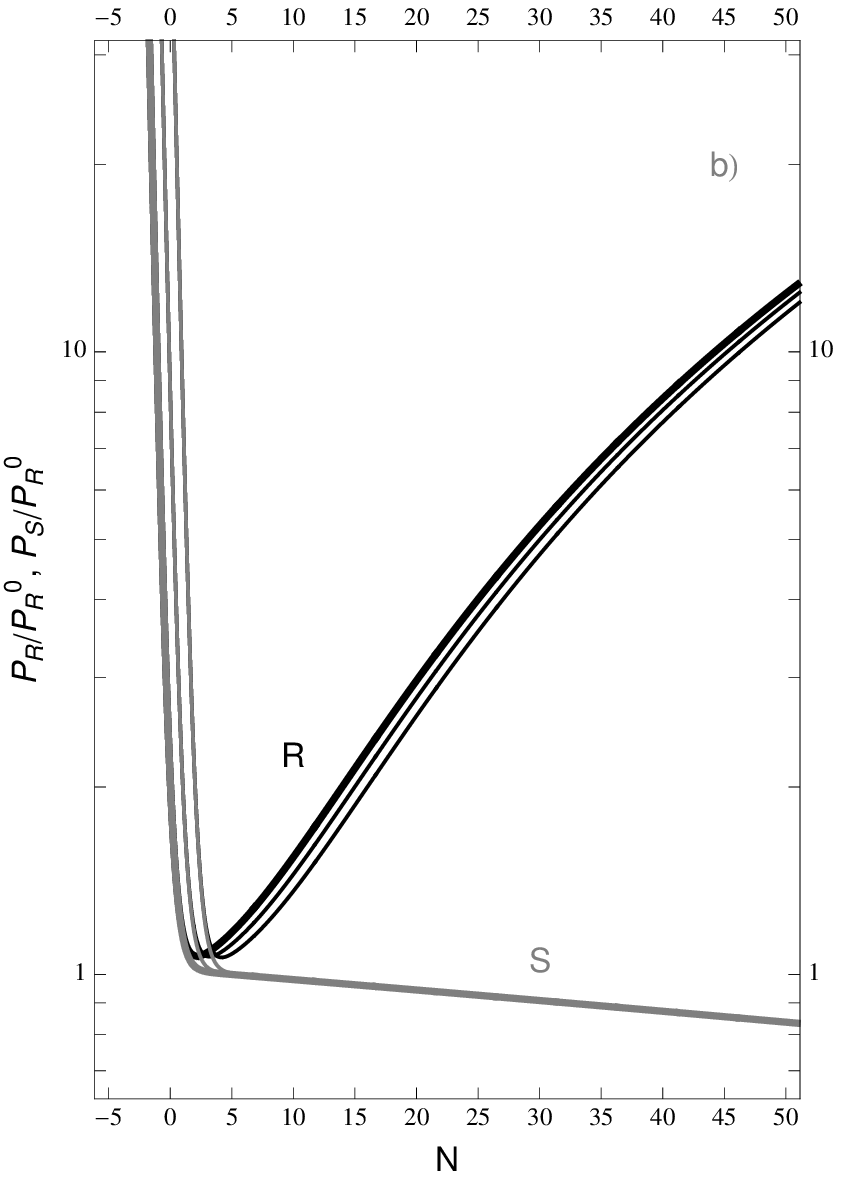}
\caption{\em (a) Evolution of the instantaneous curvature and isocurvature perturbations
shown in terms of the power spectra. Initial conditions are imposed 8 efolds before the
Hubble radius exit at $N=0$. Solid lines $\mathcal{R}$ and $\mathcal{S}$ show the total curvature and isocurvature perturbations. Dashed lines $\mathcal{R_S}$ ($\mathcal{S_R}$) correspond to the components of the curvature (isocurvature) perturbations generated from initial pure isocurvature (curvature) perturbations.  (b) Comparison of the curvature ($\mathcal{R}$) and isocurvature ($\mathcal{S}$)
modes with different wave numbers, leaving the Hubble radius at $N=0,1,2$. \label{f1}}
\end{figure}

We validate the discussion of our simple model by
a numerical analysis of a specific example
 of two-field inflation with noncanonical kinetic terms,
realizing the ideas presented above. We use a potential of the form (\ref{exp1})
and assume that the noncanonicality has a form $b(\phi)=-\gamma\phi/M_P$, with $\gamma$ a constant,
typical of many supergravity constructions.
According to (\ref{orel2}), there must be
a relation between the parameters of the model ($\alpha$, $\beta$ and $\gamma$) and the
initial condition $\phi_0$ for the field $\phi$, to ensure a straight trajectory in the field space.
We chose $\beta=1/30$, $\gamma=1$, $\phi_0=\chi_0=0$, $\dot\phi=0$, $\dot\chi=-e^{-2b}V_\chi/(3H)$ and we adjusted $\alpha=-3.7\cdot10^{-4}$
so that the condition (\ref{orel2}) was satisfied.
The equations of motion for the background values of the fields $\phi$ and $\chi$,
as well as for the gauge invariant curvature and isocurvature perturbations, can be solved using the techniques
described in \cite{Lalak:2007vi}. The initial conditions are specified at 8 efolds before the
largest modes relevant for the CMB temperature fluctuations leave the Hubble radius.
Since we expand the potential up to linear order in the fields, we cannot trace the
model-dependent exit from inflation, but we presume that rolling of a standard waterfall field
triggered by the inflaton field $\chi$ ends the slow-roll period.
Therefore,
we trace the evolution up to $N_\mathrm{e}=51$ efolds
after the observable modes with the smallest $k$ leave the Hubble radius.
The parameters $A$, $B$ and $D$ are practically constant during inflation -- their values
at Hubble radius exit are $A_\ast=1.1\cdot10^{-3}$, $B_\ast=0.067$ and $D_\ast=6\cdot 10^{-6}$.
We find that $n_s-1=-0.038$ and that
(\ref{indan}) gives precisely this value (we arranged the parameters so that other
contributions are negligible). The evolution of the curvature and isocurvature perturbations
for a few selected modes is shown in Figure \ref{f1}, where
we represent the perturbations by
their `instantaneous power spectra'
defined by
$\mathcal{P}_\mathcal{R}(k)\, \delta(\mathbf{k}-\mathbf{k}') \equiv \frac{k^3}{2\pi^2} \langle\mathcal{R}^\ast_{\mathbf{k}'}
\mathcal{R}_\mathbf{k}\rangle$ and analogously for $\mathcal{S}$, where the
linear perturbations are treated as Gaussian random variables.

The particular toy model we chose to analyze is, of course, very simplified. The potential is linear in the
fields and very flat, and the initial conditions for the field evolution are chosen so that the classical
trajectory in the field space is a straight line. 
Although string-theory-based models of inflation with a linear potential have been put forth \cite{msw},
such a negative contribution to the spectral index can also  arise in more generic settings.
For example,
\cite{Lalak:2007vi} studies the double quadratic potential $V(\phi,\chi)=\frac{m^2}{2}(\phi^2+\chi^2)$ with $b(\phi)=-\phi/M_P$ and
finds trajectories which are almost straight lines in the field space. Although the variations of the potential and the inflaton velocity during
inflation are sufficient to drive $n_s$ below 1, in that example the interactions between the curvature and isocurvature perturbations are
still responsible for roughly half of the deviation from unity. An important feature of that example is that the perturbations orthogonal
to the trajectory in the field space are light compared to the Hubble parameter, which allows one to achieve the balance condition (\ref{orel2}) approximately.

The potential (\ref{exp1}) employed here bears some resemblance to
models of Brans-Dicke inflation with $V=e^{-\alpha\phi}U(\chi)$
\cite{GarciaBellido:1995fz}, which also predict $n_s<1$. In these models
the evolution of the modulus $\phi$ suppresses the scale of the potential
as inflation proceeds, hence the amplitude of the curvature perturbations
leaving the Hubble radius decreases with time. In our example, the scale of the potential
remains practically constant throughout inflation and the red tilt arises solely from
the interactions between the perturbations.

The result shown in (\ref{indan}) may be viewed as particular limit of
a general expression for the spectral index, obtained in \cite{DiMarco:2005nq}
under assumptions of constant slow-roll parameters and uncorrelated curvature
and isocurvature perturbations at the Hubble radius crossing. However, our work goes
beyond this analysis in several ways. Firstly, we prove the existence of a
trajectory for which these assumptions are satisfied and which also
allows for a relatively large couplings between the curvature and the 
isocurvature perturbations. This is only possible in the presence of
noncanonicality: introducing such a coupling from interactions
in the potential, described by $\eta_{\sigma s}$, would produce a fast turn
in the trajectory and an instantaneous sourcing of the curvature perturbations
by the isocurvature ones and the $N_\mathrm{e}$ dependence would be lost.
Secondly, the balance condition (\ref{orel2}) for the inflationary trajectory
ensures that the spectrum of the curvature perturbations is red-tilded,
contrarily to what the general expression in \cite{DiMarco:2005nq} may
suggest. 

A recent analysis \cite{Kobayashi:2010fm} studied a setup similar to ours. The crucial difference
is that the authors of \cite{Kobayashi:2010fm} assumed {\em both} scalar fields slowly rolling,
{\em i.e.} neglected the term proportional to $\dot\chi^2$ in (\ref{bk1}) as a higher order correction.
This leads to different inflationary trajectories and thus to different predictions.
Our class of trajectories, given by (\ref{orel2}), has been previously studied in the literature only in
the `gelaton' scenario \cite{tolley}.
In \cite{tolley} it was, however, assumed that the parameter $B$ is very large and that
the field $\phi$ is much heavier than the Hubble scale. The expansion in the coupling
between the perturbations used here is insufficient in analyzing such models and we
defer a detailed comparison between differences in predictions of our example and those
of the `gelaton' scenario to future work \cite{Newpaper}.

Finally, we note that in models in which the curvature perturbations are generated from
isocurvature perturbations {\em during} inflation, the power spectrum of the curvature
perturbations is larger than the single field result (\ref{one}).
In our simple example, this enhancement is
by a factor $B^2 N_\mathrm{e}^2$, which may serve for
decoupling the scale of the inflationary potential from the amplitude of the density
perturbations
by means other than lowering the slow-roll parameter $\epsilon$.

In conclusion, in this note we have studied the inflationary dynamics of a model with two scalar fields, one with a
non-canonical kinetic term, and a very flat potential.
We found, analytically and numerically, that the interactions between the curvature and isocurvature perturbations
can give a negative contributions to $n_s$ of the order of a few percent,
accounting for practically {\em all} the redness of the spectrum.
This offers a new principle for constructing phenomenologically viable models of inflations
employing very flat potentials.


\noindent
{\it Acknowledgments:}
We thank D.~Langlois for discussions at earlier stages of this project and S. Watson for many useful comments.
This work was partially supported by the EC 6th Framework Programme MRTN-CT-2006-035863
and by TOK Project MTKD-CT-2005-029466.
The work of S.C. has been supported by the Cambridge-Mitchell Collaboration in Theoretical
Cosmology, and the Mitchell Family Foundation.
The work of Z.L.~was partly supported in part by the National Science Foundation under Grant No. PHY05-51164.
K.T.~acknowledges support from Foundation for Polish Science through its programme
Homing. K.T.~is grateful to DAMTP, University of Cambridge for hospitality and stimulating atmosphere.

\end{document}